\documentclass[twocolumn,twoside,slac_two]{revtex4}
\usepackage{graphicx}
\usepackage{fancyhdr}
\newcommand{\fermi}{\emph{Fermi}~}

\begin{document}

\title{Three Years of Fermi LAT Flare Advocate Activity}

%

\author{Stefano Ciprini, Dario Gasparrini}
\affiliation{ASI Science Data Center, Frascati, Roma, Italy}%
\affiliation{INAF Observatory of Rome, Monte Porzio Catone, Roma, Italy}%
\author{Denis Bastieri}
\affiliation{University of Padova, Padova, Italy}%
\affiliation{INFN Padova Section, Padova, Italy}%
\author{James Chiang}
\affiliation{SLAC National Accelerator Laboratory, Menlo Park, CA, USA}%

\author{Gino Tosti}
\affiliation{SLAC National Accelerator Laboratory, Menlo Park, CA, USA}%
\affiliation{University of Perugia, Perugia, Italy}%
\affiliation{University of Perugia, Perugia, Italy}%
\affiliation{\emph{(on behalf of the Fermi LAT collaboration)}.}

\begin{abstract}
The Fermi Flare Advocate (also known as Gamma-ray Sky Watcher, FA-GSW)
service provides for a daily quicklook analysis and review of the
high-energy gamma-ray sky seen by the Fermi Gamma-ray Space Telescope.
The duty offers alerts for potentially new gamma-ray sources, interesting
transients and relevant flares. A public weekly digest containing the main highlights about the GeV gamma-ray sky is published in the web-based Fermi Sky Blog.
During the first 3 years of all-sky survey, more than 150 Astronomical
Telegrams, several alerts to the TeV Cherenkov telescopes, and targets of
opportunity to Swift and other observatories have been distributed. This
increased the rate of simultaneous multi-frequency observing campaigns
and the level of international cooperation. Many gamma-ray flares from
blazars (like the extraordinary outbursts of 3C 454.3, intense flares of PKS
1510-089, 4C 21.35, PKS 1830-211, AO 0235+164, PKS 1502+106, 3C 279, 3C
273, PKS 1622-253), short/long flux duty cycles, unidentified transients
near the Galactic plane (like J0910-5041, J0109+6134, the Galactic center
region), flares associated to Galactic sources (like the Crab nebula, the
nova V407 Cyg, the microquasar Cyg X-3), emission of the quiet and active
sun, were observed by Fermi and communicated by FA-GSWs.
\end{abstract}

\maketitle

\thispagestyle{fancy}


\section{Introduction and scope of the FA-GSW service}

The Large Area Telescope (LAT), on board the \fermi Gamma-ray
Space Telescope \cite{atwood09}, is a pair-conversion $\gamma$-ray
telescope, sensitive to photon energies from about 20 MeV up to $>300$
GeV. The LAT consists of a tracker (two sections, front and back), a calorimeter and an
anti-coincidence system to reject the charged-particle
background. \fermi LAT, working in all-sky survey mode, is an optimal hunter for high-energy flares, transients and new gamma-ray sources, and is an unprecedented monitor of the variable $\gamma$-ray sky, thanks to the large peak effective area, wide field of view ($\approx 2.4$~sr), improved angular resolution and sensitivity.

%
\begin{figure}[t!!!]
\begin{center}
\hskip 0.0cm 
\resizebox{7.0cm}{!}{\rotatebox[]{0}{\includegraphics{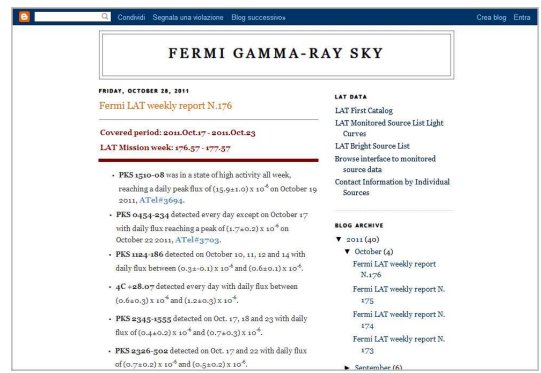}}}
\hskip 0.3cm 
\vspace*{-0.2 cm}
 \caption{The public weekly Fermi Sky Blog (web address: \texttt{\underline{fermisky.blogspot.com}}).
  ATels posted and other information about the day by day $\gamma$-ray sources detected above 100 MeV are summarized every week in this web digest. }
   \label{fig:FermiSkyBlog}
\end{center}
\vspace*{-0.8 cm}
\end{figure}
%
%

This all-sky monitoring is complemented by the Flare Advocate (a.k.a. Gamma-ray Sky Watcher, FA-GSW) duty, a scientific service belonging to the LAT Instrument Science Operations and devoted to quicklook inspection and daily review of the gamma-ray sky observed by \fermi LAT, performed with continuity for all the year through weekly shifts.

\begin{figure*}[t!!!]
\begin{center}
\hskip 0.0cm 
\resizebox{\hsize}{!}{\rotatebox[]{0}{\includegraphics{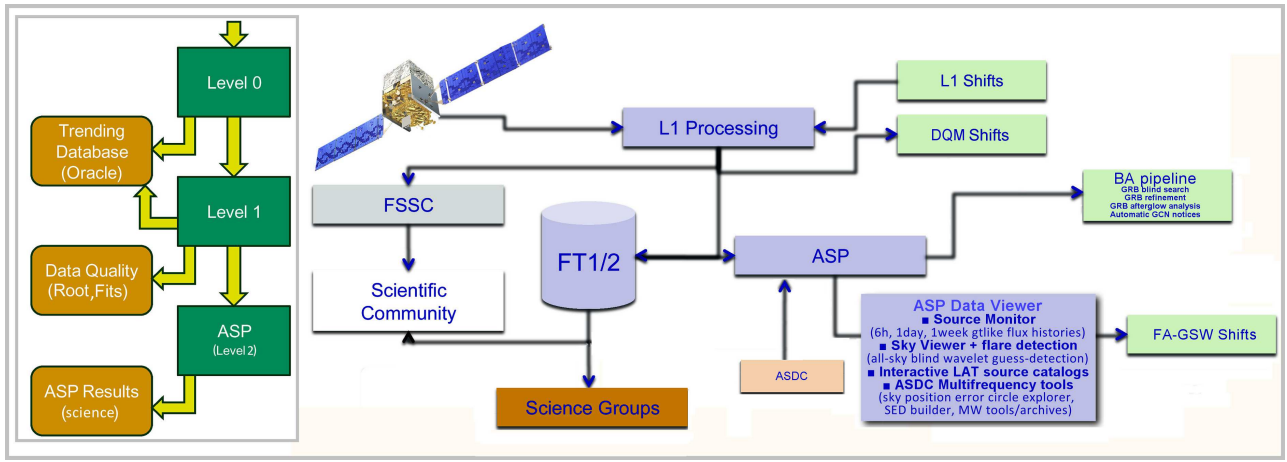}}}
\vspace*{-0.2 cm}
\caption{\textit{Left inset panel:} diagram of the main general blocks of the \fermi LAT data processing flow at ISOC (SLAC, Stanford University). The final astrophysical science data (FT1 and FT2 fits files) are sent to the Fermi Science Support Center (FSSC) of the NASA Goddard Space Flight Center
for public distribution to the scientific community.  \textit{Main panel:} block diagram of \fermi LAT data processing with final steps of duty services covered by members of the LAT collaboration in evidence (L1, Data Quality Monitor, DQM, Flare Advocate Gamma-ray Sky Watcher FA-GSW, weekly shifts). In particular the Automated Science Processing (ASP, Level 2 processing) is outlined with the links to the GRB automatic analysis pipeline and to the FA-GSW duty activity.}
   \label{fig:schemeblocks.eps}
\end{center}
\end{figure*}
%

The FA-GSW service points out basic facts and information about the $\gamma$-ray sky of potential interest for the LAT internal science groups, through a day-by-day inspection and review of the all-sky photon count maps collected and of the quicklook science pipeline results. Summaries
about the sky surveyed and monitored by \fermi LAT, transients, flaring and new sources on six-hour and 1-day time intervals are communicated along with any relevant news to the external multiwavelength (MW)
astrophysical community using the LAT-MW mailing-list\footnote{Sign up for ``gammamw'' mailing list at address:\\  \texttt{http://fermi.gsfc.nasa.gov/ssc/library/newsletter/}}. Furthermore Astronomer's Telegrams (ATels)\footnote{Web address: \texttt{http://www.astronomerstelegram.org}}, automatic burst GCNs and special GCNs for blazar flares are distributed in addiction to weekly summary reports in the ``Fermi sky Blog''\footnote{Web address: \texttt{http://fermisky.blogspot.com}} (Fig.~\ref{fig:FermiSkyBlog}). Thanks to this service joined with the public distribution of LAT data at the FSSC\footnote{Web address: \texttt{http://fermi.gsfc.nasa.gov/ssc/}}  the \fermi LAT collaboration is therefore able to promote and increase the rate of multifrequency collaborations and observations, maximizing the scientific return and rate of international scientific cooperation of the \fermi mission. First seeds for variability and MW follow-up and studies are often triggered by the FA-GSW activity (see the LAT MW Coordinating Group\footnote{Web address: \texttt{https://confluence.slac.stanford.edu/x/YQw}} and \citet{thompson09}).

\section{ASP infrastructure}

This activity is based on the automated quicklook data analysis of Level 2 (L2) at the \fermi LAT Instrument Science Operation Center (ISOC) of SLAC-Stanford (Fig. \ref{fig:schemeblocks.eps}). L2 processing (instrument monitoring pipeline, background monitoring, and quick look science analysis) is triggered by the first availability of Level 1 (L1) processed data and performed on longer time intervals (six hour, 1 day and 1 week) referred therefore as Automated Science Processing (ASP). The ASP analysis pipeline running on the final astrophysical science data (photon event files FT1, and spacecraft data files FT2 fits files) is composed of several scientific tasks (Fig. \ref{fig:schemeblocks.eps}, and \citet{chiang07,cameron07}):
%
\begin{figure}[b!!!]
\begin{center}
\hskip 0.0cm 
\resizebox{\hsize}{!}{\rotatebox[]{0}{\includegraphics{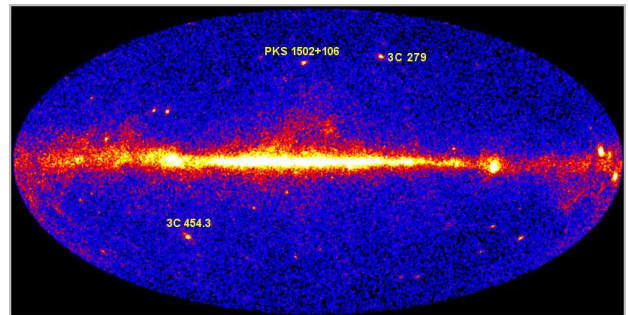}}}
\vspace*{-0.2 cm}
 \caption{The all-sky photon count map above 100 MeV accumulated in one week of \fermi LAT survey. All-sky maps like this and automatic quick-look source detections are investigated on 1-day and 6-hour time intervals day by day by Flare Advocates. 3 bright blazars of the week are labeled (3C 454.3, PKS 1502+106, 3C 279).}
   \label{fig:allskymap}
\end{center}
\end{figure}
%

\begin{table}[t!!!]
\begin{center}
\caption{The 159 ATels posted divided for topics.}
\begin{tabular}{|l|c|}
\hline \textbf{ATel type} & \textbf{Num.} 
\\
\hline %
Total Fermi ATels (on blazars mostly)	& 159\\
\hline %
Fermi on Galactic sources &	18\\
Fermi on the Sun	& 3\\
Swift results on ToOs triggered by Fermi	& 18\\
Fermi-Swift joined results	& 1\\
Fermi-Integral joined results	& 1 \\
Fermi-WEBT joined results	& 1 \\
Fermi-HESS joined results	&  1\\
\hline
\end{tabular}
\label{table:ateltypes}
\end{center}
\vspace*{-0.2 cm}
\end{table}
%
\begin{figure}[h!!!]
\vspace*{-0.3 cm}
\begin{center}
\hspace*{-0.6 cm} 
\resizebox{7.3cm}{!}{\rotatebox[]{0}{\includegraphics{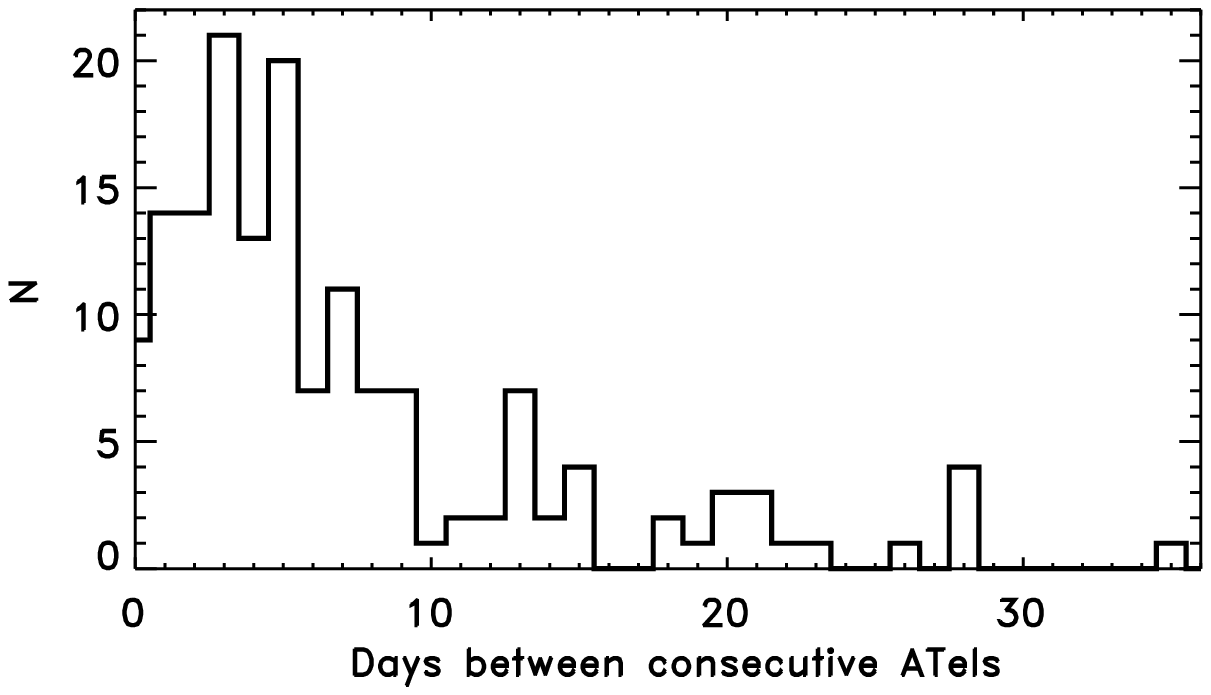}}}
\vspace*{-0.3 cm}
\hspace*{-0.6 cm} 
\resizebox{7.3cm}{!}{\rotatebox[]{0}{\includegraphics{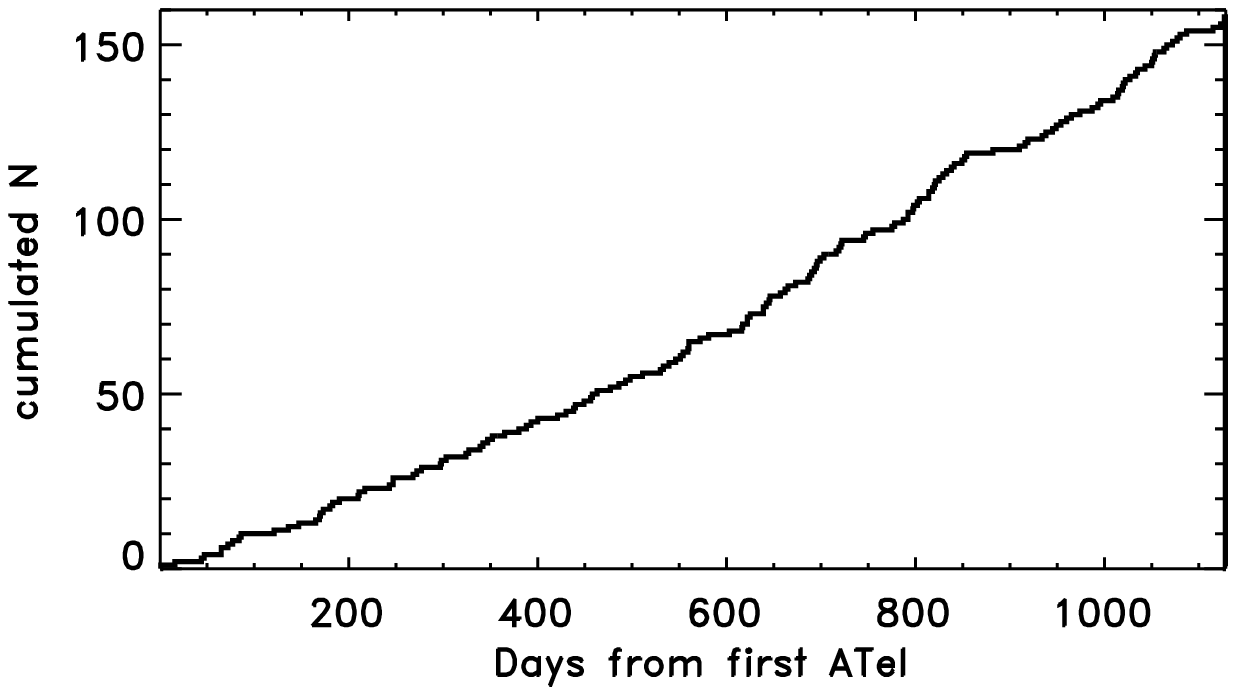}}}
 \vspace*{-0.3 cm}
 \caption{Distributions of the 159 Astronomical Telegrams (ATels) published on behalf of the Fermi LAT Collaboration from July 24, 2008 (ATel\#1628) to August 24, 2011 (ATel\#3580), i.e. in about 3 years of \fermi all-sky survey mission.
}
   \label{fig:ATELstastistics}
\end{center}
\vspace*{-0.4 cm}
\end{figure}
%
%
\begin{itemize}
  \item automatic analysis of gamma-ray bursts (impulsive transients) through refinement of parameters for LAT-detected GRBs, detection and characterization of GRBs not detected onboard, search and analysis of delayed high-energy afterglow emission;
  \item flux history monitoring based on maximum-likelihood method (gtlike science tool) of predefined list of sources (called Data Release Plan, DRP, sources) with subsequent addictions of publicly announced sources (like flaring blazars subject of ATels);
  \item blind guess-detection on all-sky photon counts maps accumulated in 6-hours, 1-day, 1-week intervals, through a fast method based on two-dimensional Mexican Hat wavelet transform, thresholding and sliding cell algorithms \citep{ciprini07};
  \item transient and flare identification based on variability test;
  \item interactive LAT source catalogs;
  \item Multi-mission/multifrequency tools and archives (like the error circle explorer and the spectral energy distribution builder) linked to ASP and provided by the ASI Science Data Center (ASDC, Roma)\footnote{Web address: \texttt{http://fermi.asdc.asi.it}}.
\end{itemize}

%

\section{Some results}

\begin{figure}[b!!!]
\begin{center}
\hskip 0.0cm 
\resizebox{6.2cm}{!}{\rotatebox[]{0}{\includegraphics{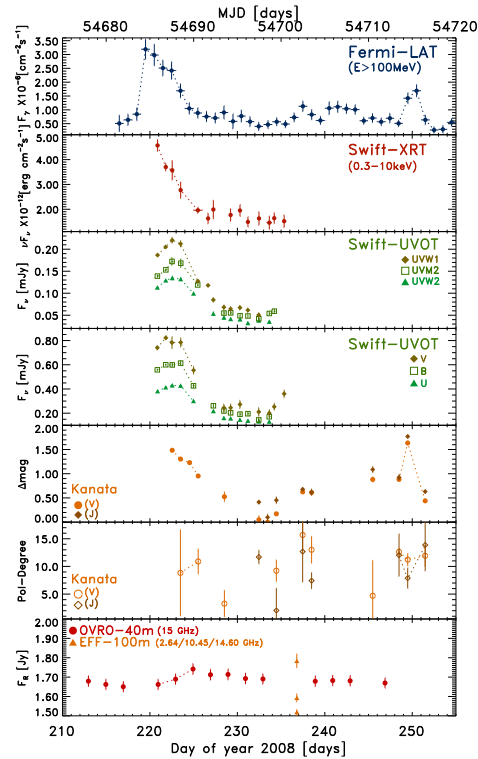}}}
\vspace*{-0.2 cm}
 \caption{Simultaneous $\gamma$-ray and multifrequency
light curves of the newly discovered $\gamma$-ray blazar PKS 1502+106 obtained during the MW campaign triggered by the outburst detected by \fermi LAT. Data reported in the panels are from \textit{Fermi}-LAT (flux above 100 MeV), \textit{Swift}-XRT (0.3-10keV flux), \textit{Swift}-UVOT (six-band optical-UV fluxes), Kanata-TRISPEC (optical-near-IR differential magnitude $\Delta V$ and $\Delta J$ bands and linear polarization), and OVRO 40m (15 GHz flux). Adapted from \citet{LAT_PKS1502}.}
   \label{fig:PKS1502}
\end{center}
\end{figure}
%

The role and activity of the FA-GSWs is therefore twofold.
\begin{itemize}
  \item Gamma-ray Flare Advocate task. Flaring sources approaching a daily flux of $10^{-6}$ photons cm$^{-2}$ s$^{-1}$ deserves attention (detection, localization, flux, photon index checked, photon counts maps and exposure maps are outlooked).
  Internal/public notes, ATels, Target of Opportunity (ToO) are submitted, MW observing campaigns are organized when needed.
  \item Gamma-ray Sky Watcher task. Results from the LAT Automated Science Processing (ASP) pipeline in 1-day and 6-hour time intervals
        are checked, searching for transients, increasing/decrasing brightness trends, and new $\gamma$-ray source candidates and spatial associations.
\end{itemize}

FA-GSWs discovered new gamma-ray blazars before the release of \fermi Catalogs, discovered several bright flares and outbursts from
blazars, some transient from low galactic latitude source, observed the emission of the quiet-sun and the flaring-sun emission.
In multifrequency science FA-GSWs triggered several targets of opportunity (ToOs) with the \textit{Swift} satellite (about a dozen per year) and involved the radio-astronomy community in joint observing programs. MW observing campaigns on several blazar and galactic source targets were also performed.

\begin{figure}[t!!!]
\vspace*{-1.0 cm}
\begin{center}
\hskip 0.0cm 
\resizebox{7.5cm}{!}{\rotatebox[]{0}{\includegraphics{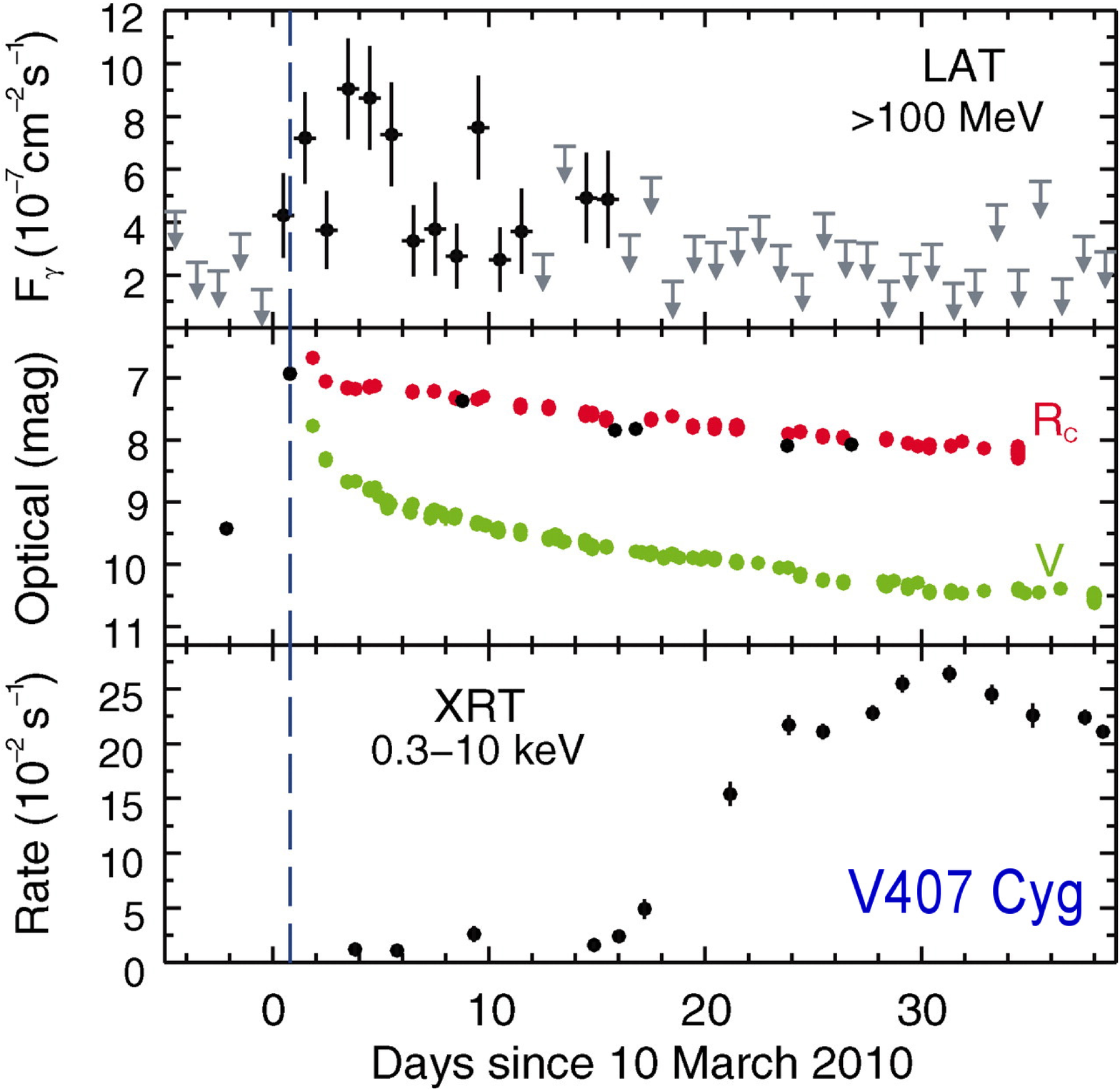}}}
\vspace*{-0.1 cm}
 \caption{Light curves of V407 Cyg in $\gamma$-rays from the Fermi-LAT (top), optical (middle), and X-rays from Swift (bottom). Adapted from \citet{LAT_V407Cyg}. }
   \label{fig:V407Cyg}
\end{center}
\vspace*{-0.3 cm}
\end{figure}
%
%

\begin{figure}[t!!!]
 \vspace*{-0.1 cm}
\begin{center}
\hskip 0.0cm 
\resizebox{\hsize}{!}{\rotatebox[]{0}{\includegraphics{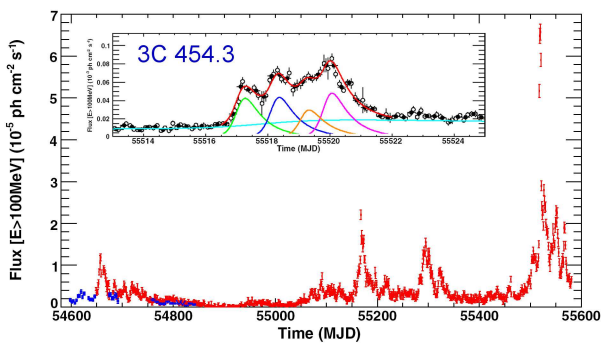}}}
\vspace*{-0.3 cm}
 \caption{Main panel daily light curve of 3C454.3 measured with the Fermi-LAT since launch. Inset panel: Light curve of the flux above 100 MeV. Open and filled symbols are 3 hr and 6 hr averaged fluxes. Five component fit (sub-flares) is outlined. Adapted from \citet{LAT_3C454_ter}.}
   \label{fig:3C454}
\end{center}
\end{figure}
%
%

In Table \ref{table:ateltypes} and Fig.\ref{fig:ATELstastistics} basic statistics about the 159 Astronomical Telegrams (ATels) published on behalf of the Fermi LAT Collaboration from July 24, 2008 (ATel\#1628) to August 24, 2011 (ATel\#3580) are illustrated. The average rate of published ATels is about one per week/shift.

More in detail the substantial menu of discoveries triggered by the FA-GSW service is: many flares from $\gamma$-ray blazars (the extraordinary outbursts of 3C 454.3 Fig. \ref{fig:3C454} and \citet{LAT_3C454_ter} large flares of PKS 1510-089, 4C 21.35, PKS 1830-211, AO 0235+164, PKS 1502+106, Fig. \ref{fig:PKS1502}, 3C 279, 3C 273, PKS 1622-253, 3C 66A, etc.); short/long activity duty cycles of bright $\gamma$-ray blazars; unidentified transients near the Galactic plane (like J0910-5041, J0109+6134, Galactic center region) or associated to Galactic sources (like the Crab nebula, the nova V407 Cyg Fig. \ref{fig:V407Cyg} and \citet{LAT_V407Cyg}, the microquasar Cyg X-3, the binary star system 1FGL J1018.6-5856), intense MeV emission from the quiet and active sun.

The all-sky variability monitor of \fermi and the continuous day-by-day service performed by FA-GSWs represents the liaison between the Fermi LAT Collaboration and the MW astrophysical/astroparticle community, always invited to observe Fermi LAT sources and to propose MW collaborations.

\begin{acknowledgments}
\footnotesize{The \fermi LAT Collaboration acknowledges support from a number of agencies and institutes for both development and the operation of the LAT as well as scientific data analysis. These include NASA and DOE in the United States, CEA/Irfu and IN2P3/CNRS in France, ASI and INFN in Italy, MEXT, KEK, and JAXA in Japan, and the K.~A.~Wallenberg Foundation, the Swedish Research Council and the National Space Board in Sweden. Additional support from INAF in Italy and CNES in France for science analysis during the operations phase is also gratefully acknowledged.}
\end{acknowledgments}

\bigskip 

\begin{thebibliography}{9}   





\bibitem[Abdo et al. (2010a)]{LAT_V407Cyg} Abdo, A.~A., Ackermann,
M., Ajello, M., et al.\ 2010a, \textit{Science}, 329, 817

\bibitem[Abdo et al. (2010b)]{LAT_PKS1502} Abdo, A.~A., Ackermann,
M., Ajello, M., et al.\ 2010b, \textit{ApJ}, 710, 810

\bibitem[Abdo et al.(2011)]{LAT_3C454_ter} Abdo, A.~A., Ackermann,
M., Ajello, M., et al.\ 2011, \textit{ApJ Lett}, 733, L26

\bibitem[Atwood et al. (2009)]{atwood09} Atwood, W.~B., Abdo, A. A., Ackermann, M., et al. 2009, \textit{ApJ}, 697, 1071

\bibitem[Cameron(2007)]{cameron07} Cameron, R.~A.\ 2007, \textit{AIP Conf Proc.}, 921, 534

\bibitem[Chiang(2007)]{chiang07} Chiang, J.\ 2007, \textit{AIP Conf. Proc.} 906, 11

\bibitem[Ciprini et al.(2007)]{ciprini07} Ciprini, S., Tosti, G.,
Marcucci, F., et al.\ 2007, \textit{AIP Conf. Proc.}, 921, 546

\bibitem[Thompson, D. J.(2009)]{thompson09} Thompson, D. J., 2009, \textit{eConf}, C0911022, (arXiv:0912.5320)

\end{thebibliography}

\end{document}